# Searching for Life-As-We-Don't-Know-It: Mission-relevant Application of Assembly Theory for Exoplanet Life Detection

***PI*: Sara Walker (sara.i.walker@asu.edu); *Co-Is:* Estelle Janin (ejanin@asu.edu), Evgenya Shkolnik (shkolnik@asu.edu), Louie Slocombe (lslocomb@asu.edu), Leroy Cronin (lee.cronin@glasgow.ac.uk)**

**Primary Topic:** Identify Emerging Themes and Technologies (2)

The history of life detection missions and biosignature science is punctured by "horizon extensions", which broadened the investigative scenery and marked critical shifts in mindset, endeavors, and research strategies [1]. This started in 1965 with Lovelock and the search for life on Mars. In the 1990s, the discovery of the first exoplanets propelled astrobiology beyond the Solar System. From 2009, Kepler revealed an unexpected diversity of exoplanets and stellar systems, shifting our likelihoods away from Sun-twin stars and Earth-twin planets and encouraging re-evaluations of biosignatures in different scenarios. However, even though the properties we associate with life and the environments we seek to discover it have both changed tremendously over 80 years, biosignature concepts themselves have remained relatively static with few major innovations. In the case of exoplanets, biosignature science is still mostly focused on the spectroscopic detection of gases in chemical disequilibrium (especially $O_2$ and $CH_4$) and the detection of Earth-like life in Earth-like environments, rooted in a lineage of ideas stemming from those first proposed by Lovelock in 1965 [2] [3]. Notably, the past decade has witnessed substantial effort to integrate biosignature concepts into various detection and validation protocols to establish appropriate confidence thresholds for remote life detection [4]. Meanwhile, the list of potential mechanisms capable of generating false positives for conventional biosignature gases is growing faster than the putative ways of detecting life on exoplanets [5]. This has led to an increasing dependence of these biosignature signals on a large amount of contextual information about the planet and system at large, which is believed necessary to rule out abiotic scenarios [6]. It is arguable whether these can be exhaustively searched. Taken together, these realizations warrant a pressing need for conceptual innovation and complementary approaches, consistent with the growing diversity of astrobiology targets and balancing the need to contextualize the system in unfeasible detail.

In recent years, there has been a growing interest in adapting quantitative tools from complex systems science to characterize exoplanets and to conceptualize new biosignatures. This trend reflects a recognition that exoplanet astrobiology involves complex, multivariate and high-dimensional data that challenge traditional strategies. Examples include quantifying the statistical complexity of planetary features using epsilon machines and assessing the temporal variability of light reflectance spectra [7], developing an information-entropic method to decompose planetary spectra and identify biosignature pairs at stages of Earth's evolution [8], or using network-based approaches to characterize the atmospheric topology of inhabited planets [9] [10]. However, complexity-based approaches have been primarily exploratory and not grounded in measurement. To significantly advance the field, agnostic life detection methods should address two critical points: (1) they should allow testing hypotheses about the generalized nature of life and how it shapes planetary atmospheres, and (2) they should be implementable in astronomical datasets, with a clear understanding of observational relevance for future missions.

A biosignature framework that bridges these two requirements is Assembly Theory (AT), which if adapted to the study of planetary atmospheres, could offer new approaches to exoplanet life detection that are observationally accessible. AT offers an opportunity to advance biosignature science by connecting its theoretical, experimental, and observational components by quantifying molecular complexity as the minimum number of recursive steps required to combinatorically assemble observed molecular objects from a pool of building blocks (the Assembly Index, or AI). The theory has been shown to capture signatures of selection and evolution in a variety of systems [11] and to accurately identify unambiguous molecular biosignatures – with measurement of AI in organic molecules empirically validated through nuclear magnetic resonance, tandem mass spectrometry and infrared spectroscopy [12] [13]. Importantly, this framework is agnostic to life's specific instantiation and unique among the developing field of agnostic approaches in being amenable to direct measurement. It leverages fundamental and quantifiable characteristics of evolutionary systems – i.e. hierarchical modularity, recursivity, historical contingency, and selection – to provide a robust and falsifiable theoretical framework for chemical data interpretation and life detection, and a means to test it observationally and experimentally.

An AT-based biosignature framework tests the hypothesis that complexity generated at a planetary scale by a global biosphere/technosphere translates into the atmosphere and influences measurable properties of its chemical space. The systems we study are graph-theoretic representations of atmospheric composition at a given time, i.e., ensembles of graphs corresponding to the chemical species in the atmosphere, detected above a given abundance threshold. For each atmosphere, we calculate the minimum number of steps it takes to co-construct all the molecules from the pool of available chemical bonds, along the shortest combinatorial pathway. Unlike other network-based approaches [9] [10], AT does not rely on knowledge of kinetics, meaning that it can be applied with fewer assumptions than other complexity-based approaches. This analysis can be applied for various timescales (e.g., different planetary evolution stages), and system-level characteristics (e.g., secondary or primary atmosphere, ocean-covered worlds, orbiting Sun-like stars or M-dwarfs) – thus building up an understanding of how planetary atmospheres behave and cluster in terms of their Assembly outputs, and how much evolution and selection is evident in their chemical structure. It remains to be verified, but this also opens the possibility of determining a threshold in atmospheric complexity necessary for definitive claims of life, like the threshold for definitive claims in organic molecular assembly identified by Marshall *et al.* [12].

So far, comparing Solar System atmospheres and exoplanet archetypes like Lava Worlds or Ocean Worlds, we find that the Earth's atmosphere displays the largest complexity, independent of any observational bias (e.g., by imposing a detectable abundance cutoff) [14]. Earth's atmospheric chemical space is also consistent with a deeper "co-construction" of molecular species, with higher reuse of chemical bonds and intermediate chemical fragments – which we anticipate all complex, inhabited worlds to exhibit. Species in the Earth's atmosphere also reflect a more exhaustive exploration and "actualization" of the possibility space for atmospheric chemistry, which is constrained by the nature and number of available bonds. For example, Venus and Earth have relatively the same diversity of bonds yet the Earth has more diversity in the number of molecules above a given abundance cutoff. This suggests that something about the Earth as a planetary system (potentially life) allows more molecular diversity in its atmosphere than Venus, even though similar bonds are stable. In this context, Venus is a higher entropy world, making most chemical species without selectivity. This is particularly insightful because it highlights the usefulness of this approach for other open problems in exoplanet science beyond life detection. It could provide new methods to infer the presence of undetected species from atmospheric observations – currently a significant bottleneck in conventional spectral retrieval pipelines.

The AT-based framework aims at delivering a classification scheme of planetary atmospheres, not based on their composition, but instead on how easy (abiotic) or hard (biotic) it is to co-construct molecules within their atmosphere (i.e., how much selection was necessary [11]). Therefore, it holds the potential to provide a quantitative approach to assess how much selection and evolution at a planetary scale is required to generate an observed composition, within a strategy that is not restricted to life-as-we-know-it and is comparable across a wide variety of planets. Applying AT to planetary atmospheres addresses some of exoplanet astrobiology's deepest challenges, by resting on a firm hypothesis about the nature of life and collapsing the dimensionality of the exoplanet problem to focus on the specific features of the system relevant to a well-defined, falsifiable hypothesis (i.e., that life generates high assembly configurations of matter) for life detection.

**Secondary Topic:** Evaluate Astrobiology's Role in Missions (6)

Establishing practical, alternative approaches to life detection and renewed perspectives on ubiquitous concepts like complexity, evolution, and habitability demands a tight connection to missions and observations from the early stages of a framework's development. This highlights the benefits of co-developing a novel theoretical framework with the ongoing planning of near-term and future observations, such as the Habitable Worlds Observatory (HWO). Selected as NASA's next flagship mission by the Astro2020 Decadal Survey, HWO will be the first telescope designed explicitly for life detection on exoplanets [15]. So far, HWO development has been primarily aligned with detecting Earth-analog worlds. While this consensus approach provides a solid foundation, the revolutionary nature of HWO could significantly benefit from more agnostic strategies that look beyond Earth-based assumptions. In general, astrobiology missions must be compatible with a wide range of complementary approaches that target similar observational criteria, maximizing our potential to detect both Earth-like life and life-as-we-don't-know-it. Applying AT to atmospheres and generating simulated observations consistent with different sets of HWO instrumental parameters and noise constraints allows quantitatively informing technical aspects of the mission as early as the pre-Phase A stage. This demands rigorously and holistically mapping trade-offs between the accuracy/precision of theoretical outputs and different instrumental requirements, especially wavelength cutoffs and spectral resolution, which strongly affect the detection of species of astrobiological interest (e.g., CO, $CO_2$, $SO_2$, $N_2O$, $NO_2$ or $NH_3$). In this context, future work should extend our initial analyses focused on a few planetary archetypes to more continuous population studies, informed by the HWO Preliminary Input Catalog [16]. Therefore, while enhancing the observatory's astrobiological potential, this work would leverage existing target plans and mission strategies of other HWO initiatives, highlighting major convergent/divergent areas to ensure efficient resource allocation and minimize redundancy.

To optimize the relevance of novel biosignature frameworks for current, near-term and future exoplanet missions, these approaches should be useful and applicable to any chemical space, from abiotic to biotic environments. This allows the framework to be iteratively improved as different observations accumulate and to provide a satisfactory empirical grounding in the early stages of data interpretation. At a time where exoplanet atmospheric data is much more readily available for larger, hotter planets like hot Jupiters or sub-Neptunes/super-Earths, the framework can thus be tested against predictions in a much shorter timescale. By offering a standardized pipeline through which to classify *any* planet according to the amount of selection present in its atmosphere, an AT-based framework can be validated by population studies sampling a wide diversity of the exoplanet parameter space, thus providing a continuum of planetary relatedness according to how much selection has operated over their history and is evident in the atmosphere's present characteristics. This allows breaking away from a binarization of systems according to a strict "alive/dead" dichotomy and instead to capture and characterize the transition between abiotic and living worlds, potentially recognizing that the distinction may not be a hard phase boundary.

Finally, the exoplanet field currently contends with the challenges associated with different biases embedded in the forward models of atmospheric spectral retrievals, which are essential to interpretating of observational data. Several intercomparison studies to quantify these discrepancies have emerged over the past few years [17]. To minimize the impact of pipeline-specific biases, a goal should be to build up from an existing correlation of Assembly with infrared data [13] and to test the application of an AT-based framework on laboratory spectra of gaseous mixtures. This will allow determining if Assembly can be directly inferred from the reduced planetary spectra without necessarily retrieving the nature of the chemical species involved. Suppose this is not possible given current or planned capabilities; in that case, it will necessitate instead determining the appropriate step to incorporate our measures in retrieval schemes, in a way that remains agnostic and possibly replaces more poorly constrained, kinetic-based inferences.

**Secondary Topic:** Foster Cross-Divisional and Cross-Directorate Collaboration (3)

Applying complex systems approaches to life detection efforts promotes collaborative research and exploration across the broad scientific and programmatic pursuits, spanning SMD divisions such as Planetary Science and Astrophysics. In the context of the Planetary Science division, it formulates novel, operational hypotheses about how life began on Earth, evolved, and can be unambiguously detected – centered around the emergence of selection and the generation of increasingly complex objects in the case of AT – to assess whether it has evolved on other planets, inside or outside our Solar System. By providing a common lens through which both Solar System planets and exoplanets can be studied, and by offering a continuum of planetary relatedness rather than binary categorizations, these approaches encourage insightful synergies between Solar System and exoplanet science, meaningfully situating the former in the broader context of the latter. These strategies are also relevant to Astrophysics as they seek to answer one of the major questions highlighted in this division: "Are we alone?" and "Does life, be it similar to our own or not, exist elsewhere?". Developing new frameworks that are both empirically grounded and mission-oriented, aligns with NASA's current discussion of a "future mission concept with the capabilities to identify and image Earth-like planets and characterize them for evidence of life", such as the Habitable Worlds Observatory selected as NASA's next flagship mission by the Astro2020 Decadal Survey [15].

It is imperative to encourage the development of exoplanet biosignature frameworks that span the theoretical, experimental and observational aspects of biosignature science, ensuring tight feedback loops between the relevant actors and enabling to test predictions across a wider diversity of interconnected hypotheses. So far, AT has demonstrated its use for biosignature research (both remote and in-situ), origins-of-life studies, and synthetic biology efforts, thus informing and connecting many of astrobiology's core tenets – understanding the origin, evolution, distribution, and future of life in the Universe, within a conceptual framework with the potential to unify previously disparate approaches. By agnostically integrating insights across time and scales, complex system approaches can uniquely support the evolving needs of a growing and increasingly interdisciplinary astrobiology community, while rigorously enforcing falsifiable and testable predictions.

Finally, to optimize life detection efforts, it is also essential to identify the assumptions and hypotheses underlying each strategy. For example, focusing on $O_2$ as a biosignature strongly selects for complex life based on Earth-like photosynthesis, and relies on the enumeration of all possible false positive scenarios to rule out an abiotic explanation. In contrast, an AT-based framework is agnostic to any specific metabolism or instantiation, with the limitation that it only applies to the discovery of global biospheres/technospheres whose complexity has translated into the atmosphere over evolutionary timescales. In addition, while it precludes false positive detections, it might allow false negatives. This underscores the importance of identifying and recognizing limitations in exploring possibilities and integrating diverse, complementary frameworks toward the same goal.